\documentclass[aps,pre,preprint, showpacs]{revtex4}
\usepackage{setspace}
\usepackage{amsmath}
\usepackage{amssymb}
\usepackage{epsfig}
 \normalsize
\begin{document}
\preprint{}
\title{The influence of charge and flexibility on smectic phase formation in filamentous virus suspensions}
\author{Kirstin R. Purdy}\altaffiliation[current address: ]{Department of Materials Science and Engineering, University of Illinois at Urbana Champaign, Urbana, Illinois 61801}
 \affiliation{Complex Fluids Group, Martin Fisher School of
Physics, Brandeis University, Waltham, Massachusetts 02454}
\author{Seth Fraden}
\affiliation{Complex Fluids Group, Martin Fisher School of
Physics, Brandeis University, Waltham, Massachusetts 02454}
\date{\today}
\begin{abstract}
We present experimental measurements of the cholesteric-smectic
phase transition of suspensions of charged semiflexible rods as a
function of rod flexibility and surface charge. The rod particles
consist of the bacteriophage M13 and closely related mutants,
which are structurally identical to this virus, but vary either in
contour length and therefore ratio of persistence length to
contour length, or vary in surface charge. Surface charge is
altered in two ways; by changing solution pH and by comparing M13
with {\it fd} virus, a mutant which differs from M13 only by the
substitution of a single charged amino acid for a neutral one per
viral coat protein. Phase diagrams are measured as a function of
particle length, particle charge and ionic strength. The
experimental results are compared with existing theoretical
predictions for the phase behavior of flexible rods and charged
rods. In contrast to the isotropic-cholesteric transition, where
theory and experiment agree at high ionic strength, the
nematic-smectic transition exhibits complex charge and ionic
strength dependence significantly different from predicted phase
behavior. Possible explanations for these unexpected results are
discussed.
\end{abstract}
\pacs{64.70.Md, 61.30.St}
 \maketitle

\section{Introduction}
In a suspension of hard or charged rods, purely repulsive entropic
interactions are sufficient to induce liquid crystal ordering.
Theoretically, hard rods exhibit isotropic, nematic, smectic and
columnar liquid crystal phases with increasing concentration
\cite{Onsager49,Bolhuis97a,Polson97}. Unfortunately, production of
hard, rigid, monodisperse rods is very difficult. Rigid and
flexible polyelectrolyte rods, however, are abundant, especially
in biological systems, which by nature lend themselves to
mass-production. In this paper we will study the influence of
flexibility and electrostatic interactions on the formation of a
smectic phase from a nematic phase using suspensions of charged,
semiflexible {\it fd} and M13 virus rods. Viruses, such as {\it
fd}, M13, and Tobacco Mosaic Virus are a unique choice for use in
studying liquid crystal phase behavior in that they are
biologically produced to be monodisperse and are easily modified
by genetic engineering and post-expression chemical modification.
These virus particles and $\beta$-FeOOH rods are, to our
knowledge, the only colloidal systems known to exhibit the
predicted hard-rod phase progression from isotropic (I) to nematic
(N) or cholesteric and then to smectic (S) phases with increasing
rod concentration \cite{Meyer90,Fraden95,Maeda03}. Even though
qualitative theories have been developed to describe either the
effects of electrostatics or the effects of flexibility on the
nematic-smectic (N-S) phase transition of hard rods
\cite{Kramer00,Dogic97,Tkachenko96}, they have yet to be
thoroughly tested experimentally. Near the N-S transition, the
particles are at very high concentrations, and as we will show,
dilute-limit approximations of interparticle interactions, which
are appropriate at the isotropic-nematic transition, cannot be
used. By measuring the N-S transition of charged and/or flexible
rods we learn about both the influence of these parameters on
smectic phase formation and the interactions between these rods in
concentrated suspensions. Additionally, our results add insight
into the ordering of other important rodlike polyelectrolytes such
as DNA, which often appears in high concentrations under
physiological conditions and exhibits cholesteric and columnar,
but not the smectic, liquid crystalline
phases~\cite{DNA_Strey,DNA_Livolant}.

In this paper we test the limits of current theoretical
predictions for the nematic-smectic phase transition in three
ways. First, we measure the phase transition for semiflexible
filamentous virus of identical structure and varied length. By
changing the rod length and leaving local particle structure
constant, the persistence length $P$ of the rods, defined as one
half the Kuhn length, remains constant. Subsequently, the rod
flexibility, as defined by the ratio of persistence length to
contour length $L$, or $P/L$, is altered. In our experiments the
flexibility of the particles remains within the semiflexible
limit, meaning $P\sim L$. Altering the particle flexibility within
the semiflexible limit probes the competition between rigid and
flexible rod phase behavior. Second, we vary the ionic strength of
the virus rod suspensions allowing us to probe the efficacy of
theoretical approximations for incorporating electrostatic repulsion into
hard-particle theories. Third, we measure the nematic-smectic
phase transition for filamentous virus of different charge.
Altering the surface charge by two independent techniques,
solution chemistry and surface chemistry, probes the importance of
the details of the surface charge distribution in determining long
range interparticle interactions. By varying these three
independent variables, length, charge and solution ionic strength,
we systematically examine how electrostatic interactions and
flexibility experimentally effect the nematic-smectic phase
boundary.

The colloidal rods we use are the rodlike semiflexible
bacteriophages {\it fd} and M13 which form isotropic (I),
cholesteric (nematic) and smectic (S) phases in solution with
increasing virus concentration
\cite{Lapointe73,Tang95,Dogic97,Dogic01}. The free energy
difference between the cholesteric and nematic (N) phases is
small, and therefore it is appropriate to compare our results with
predictions for the nematic phase \cite{deGennes93}. Furthermore,
near the nematic-smectic transition, the cholesteric unwinds into
a nematic phase \cite{Dogic00c}. M13 and {\it fd} are composed of
2700 major coat proteins helicoidally wrapped about the single
stranded viral DNA. They differ from one another by only one amino
acid per major coat protein; the negatively charged aspartate
(asp$_{12}$) in {\it fd} is substituted for the neutral asparagine
(asn$_{12}$) in M13 \cite{Marvin94}. They are thus ideal for use
in studying the charge dependence of the virus rod phase
transitions. Changes in the surface charge of the particles were
also achieved by varying the pH of the solution
\cite{Zimmermann86}. Additionally, by varying the length of the
M13 DNA we created M13 mutants which differ only in contour
length. The M13 mutants have the same local structure, and thus we
assume persistence length, as M13. These mutant M13 viruses were
used to measure the flexibility dependence of the nematic-smectic
phase transition.

\section{Electrostatic Interactions}
\begin{figure}
\centerline{\epsfig{file=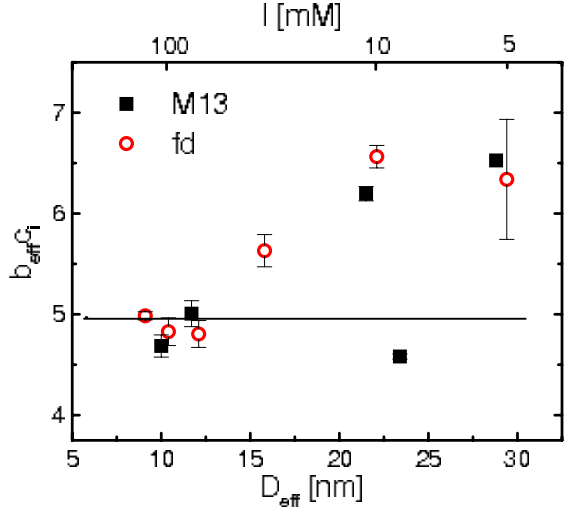,width=7.cm}}\caption[]{\label{beffci.fig}
(Color online) Isotropic-Nematic phase transition
$b_{\mbox{\scriptsize eff}}c_i$ plotted as a function of
$D_{\mbox{\scriptsize eff}}$ for both M13 and fd suspensions in
Tris-HCl buffer at pH 8.2. The original data for this figure was
published previously~\cite{Purdy04}. The solid line is the
hard-rod prediction for semiflexible rods with a persistence
length of 2.2 $\mu$m\cite{Chen93}. For small values of
$D_{\mbox{\scriptsize eff}}$ (high ionic strength), the
coexistence concentrations for the charged rods are effectively
mapped to the hard-rod predictions. The ionic strength scale is
for fd suspensions (M13 has a lower surface charge, thus
$D_{\mbox{\scriptsize eff}}$ at the same ionic strength is
slightly larger).}
\end{figure}

For colloidal rods, the total rod-rod interparticle interaction
includes a combination of hard core repulsion and long ranged
electrostatic repulsion. We present here two ways which have been
previously proposed for incorporating electrostatic interactions
into hard-rod theories for the nematic-smectic phase transition.
The first originates from Onsager's calculation of an effective
hard-core diameter ($D_{\mbox{\scriptsize{eff}}}$) which is larger
than the bare diameter $D$. $D_{\mbox{\scriptsize{eff}}}$ is
calculated from the second virial coefficient of the free energy
for charged rods in the isotropic phase \cite{Onsager49}.
Specifically, for hard, rigid, rodlike particles, the limit of
stability of the isotropic phase against a nematic phase is given
by the Onsager relation $bc_i=4$ where $b=\pi L^2 D/4$ and $c_i$
is the isotropic number density of rods \cite{Onsager49}. For
charged particles, Onsager showed that the stability condition
remains unchanged provided $D$ is replaced with
$D_{\mbox{\scriptsize eff}}$, thus $b_{\mbox{\scriptsize eff}}
c_i=4$, with $b_{\mbox{\scriptsize eff}}=\pi L^2
D_{\mbox{\scriptsize eff}}/4$. Increasing ionic strength decreases
$D_{\mbox{\scriptsize{eff}}}$, and for highly charged colloids,
like M13 and {\it fd}, $D_{\mbox{\scriptsize{eff}}}$ is nearly
independent of surface charge due to the non-linear nature of the
Poisson-Boltzmann equation, which leads to counterion condensation
near the colloid surface \cite{Tang95}. In previous work
\cite{Tang95,Purdy04}, the prediction that $b_{\mbox{\scriptsize
eff}} c_i$ is constant at the I-N phase boundary has been
experimentally verified at high ionic strength (I$>$ 60 mM, large
$L/D_{\mbox{\scriptsize eff}}$) for our system of virus rods, as
illustrated in Fig. \ref{beffci.fig}. These results validated the
mapping of the I-N transition of charged virus rods at high ionic
strength onto a hard rod theory by using an effective hard
diameter, $D_{\mbox{\scriptsize eff}}$. At low ionic strength
(I$<$ 60 mM), the prediction that $b_{\mbox{\scriptsize eff}} c_i$
is constant did not hold because of the breakdown of the second
virial approximation at small $L/D_{\mbox{\scriptsize eff}}$
\cite{Purdy04}. Consequently, we expect that if
$D_{\mbox{\scriptsize eff}}$ works to describe the N-S transition
of virus rods it would most likely be at high ionic strengths.

Since $D_{\mbox{\scriptsize eff}}$ is valid only where the second
virial approximation is valid, ie. in the isotropic phase,
Stroobants {\it et. al} developed an approximate way to describe
the electrostatic interactions in the nematic phase using this
second virial approximation \cite{Stroobants86}. They defined a
nematic effective diameter
$D^{\mbox{\scriptsize{N}}}_{\mbox{\scriptsize{eff}}}$, which is
calculated from the isotropic effective diameter
$D_{\mbox{\scriptsize{eff}}}$:
$D^{\mbox{\scriptsize{N}}}_{\mbox{\scriptsize{eff}}}=
D_{\mbox{\scriptsize{eff}}}[1+h \eta (f)/\rho (f)]$, where
\begin{eqnarray}\rho (f)= \frac{4}{\pi}\langle
\langle \sin{\phi}\rangle \rangle \end{eqnarray} \noindent and
\begin{eqnarray}
\eta (f)=\frac{4}{\pi}\langle
\langle-\sin{\phi}\log(\sin{\phi})\rangle \rangle
-(\log(2)-1/2)\rho(f)
\end{eqnarray}

\noindent The average $\langle...\rangle$ is over the solid angle
$\Omega$ weighted by the nematic angular distribution function
$f(\Omega)$ with $\phi$ describing the angle between adjacent rods
\cite{Stroobants86a}. The parameter $h=\kappa
^{-1}/D_{\mbox{\scriptsize{eff}}}$, where $\kappa^{-1}$ is the
Debye screening length, characterizes the preference of charged
rods for twisting. Crossed charged rods have a lower energy than
parallel charged rods, and $h$ correspondingly increases with
increasing electrostatic interactions (decreasing ionic strength).
This definition for the nematic effective diameter is accurate as
long as the average angle between the rods and the nematic
director $\sqrt{\langle \theta^2 \rangle}$ is much greater than
$D^{\mbox{\scriptsize{N}}}_{\mbox{\scriptsize{eff}}}/L$
\cite{Vroege92}. In this limit, the second virial coefficient is
still much larger than the higher virial coefficients, which can
be neglected. Near the N-S transition the order parameter, as
determined by x-ray measurements of magnetically aligned samples,
is $S=0.94$ \cite{Purdy03}. Using an angular distribution function
with this order parameter of $S=0.94$ we find that
$D^{\mbox{\scriptsize{N}}}_{\mbox{\scriptsize{eff}}}=1.16D_{\mbox{\scriptsize{eff}}}$
at 5mM ionic strength and
$D^{\mbox{\scriptsize{N}}}_{\mbox{\scriptsize{eff}}}=1.10D_{\mbox{\scriptsize{eff}}}$
at 150 mM ionic strength. This corresponds to $\sqrt{\langle
\theta^2
\rangle}\sim6D^{\mbox{\scriptsize{N}}}_{\mbox{\scriptsize{eff}}}/L$
for the largest value of
$D^{\mbox{\scriptsize{N}}}_{\mbox{\scriptsize{eff}}}$.

Previously, we argued that
$D_{\mbox{\scriptsize{eff}}}^{\mbox{\scriptsize{N}}} $, which is
independent of virus concentration, could describe the
electrostatic interactions of {\it fd} virus suspensions at the
nematic-smectic transition \cite{Dogic97}. However, as mentioned
above, the use of $D_{\mbox{\scriptsize eff}}$ beyond the regime
where the second virial coefficient quantitatively describes the
system is not justified, and from our calculation of
$\sqrt{\langle \theta^2 \rangle}$ we know the use of the second
virial approximation is questionable. In this article, our
expanded range of measurements of the N-S transition of virus
suspensions as a function of ionic strength, virus length, and
virus surface charge will demonstrate that electrostatic
interactions at the nematic-smectic transition are much more
complex than those predicted at the limit of the second virial
coefficient.

An alternative method for incorporating electrostatics into a
hard-rod theory for the N-S transition was developed by Kramer and
Herzfeld. They calculate an ``avoidance diameter"
$D_{\mbox{\scriptsize{a}}}$ which minimizes the scaled particle
expression for the free energy of charged parallel spherocylinders
as a function of concentration \cite{Kramer00}. With respect to
ionic strength, $D_{\mbox{\scriptsize{a}}}$ exhibits the same
trend as $D_{\mbox{\scriptsize{eff}}}$, but unlike
$D_{\mbox{\scriptsize{eff}}}$, $D_{\mbox{\scriptsize{a}}}$ is
inherently concentration dependent, decreasing with increasing rod
concentration. $D_{\mbox{\scriptsize{a}}}$ can never be greater
than the actual rod separation, something which is not impossible
with $D_{\mbox{\scriptsize{eff}}}$. Furthermore, by using the
scaled particle theory, third and higher virial coefficients are
accounted for in an approximate way \cite{Cotter78,Cotter79},
unlike Onsager's effective diameter. This makes the ``avoidance
diameter" more appropriate for incorporating electrostatic
interactions at the nematic-smectic transition. One disadvantage
with the free energy expression developed by Kramer and Herzfeld
is that it does not reduce to Onsager's theory in the absence of
electrostatic interactions. Nevertheless, Kramer and Herzfeld's
calculations do qualitatively reproduce previously published data
for the N-S transition of {\it fd} virus \cite{Kramer00}.
However, the limited range of data previously available did
not include some of the interesting features described in this
theory, which we are now able to test.


\section{Materials and Methods}
Properties of the wild type ({\it wt}) virus {\it fd} and M13 include their
length $L=$0.88 $\mu$m, diameter $D=6.6$ nm, and persistence
length $P=$2.2 $\mu$m \cite{Fraden95}. The M13 mutants have the
same diameter as the wild type M13 and lengths of 1.2 $\mu$m, 0.64
$\mu$m, and 0.39 $\mu$m \cite{Dogic01}. Because the molecular
weight of the virus is proportional to its length, the molecular
weight of the M13 mutants is $M=M_{wt} L/L_{wt}$, with
$M_{wt}=1.64\times 10^7$ g/mol and $L_{wt}= 0.88$ $\mu$m, the
molecular weight and length of wild type M13, respectively. Virus
production is explained elsewhere \cite{Maniatis89b}. Two of the
length-mutants (0.64 $\mu$m and 0.39 $\mu$m) were grown using the
phagemid method \cite{Maniatis89b,Dogic01}, which produces
bidisperse solutions of the phagemid and the 1.2 $\mu$m helper
phage. Sample polydispersity was checked using agarose gel
electrophoresis on the intact virus, and on the viral DNA.
Excepting the phagemid solutions which were 20\% by mass 1.2
$\mu$m helper phage, virus solutions were highly monodisperse as
indicated by sharp electrophoresis bands. All of these virus
suspensions form well defined smectic phases \cite{Dogic01}.

All samples were dialyzed against a 20 mM Tris-HCl buffer at pH
8.2 or 20 mM Sodium Acetate buffer adjusted with Acetic Acid to pH
5.2. To vary ionic strength, NaCl was added to the buffering
solution. The linear surface charge density of {\it fd} is
approximately 10 e$^-$/nm (3.4$\pm ~0.1$e$^-$/coat protein) at pH
8.2 and 7 e$^-$/nm (2.3$\pm ~0.1$e$^-$/coat protein) at pH 5.2
~\cite{Zimmermann86}. The M13 surface charge is 7e$^-$/nm
(2.4$\pm~ 0.1$e$^-$/coat protein) at pH 8.2 and  3.6 e$^-$/nm
(1.3$\pm~0.1$e$^-$/coat protein) as determined by comparing the
M13 composition and electrophoretic mobilities to those of {\it
fd} \cite{Purdy04}. On the viral surface, {\it fd} has four
negatively ionizable amino acids and one positively ionizable
amino acid per coat protein. At neutral pH, the terminal amine
contributes approximately $+1/2$e charge. The M13 surface has
three negatively ionizable amino acids, one positively ionizable
amino acid and the terminal amine ($+1/2$e) per coat protein.

\begin{figure}
\centerline{\epsfig{file=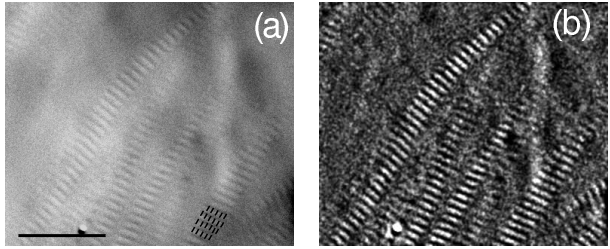,width=8.cm}}\caption[]{\label{DIC.fig}
(a) Differential interference contrast microscopy image of
nematic-smectic coexistence of {\it fd} virus suspensions. The
smectic phase can be recognized by the ladder like structures. The
virus rods are oriented perpendicular to the layers as
illustrated. The uniform texture is the nematic phase. (b)
Digitally enhanced image of (a). The scale bar is 10 $\mu$m. }
\end{figure}

After dialysis, the virus suspensions were concentrated via
ultracentrifugation at 200 000g and diluted to concentrations just
above the N-S transition. Virus suspensions were then allowed to
equilibrate to room temperature.  Bulk separation of the nematic
and smectic phases is not observed, perhaps because of the high
viscosity of the suspensions near the N-S transition. However,
smectic or nematic domains can be observed using differential
interference contrast microscopy in coexistence with predominantly
nematic and smectic bulk phases, respectively. Typically
coexistence is observed as ribbons of smectic phase reaching into
a nematic region, as shown in Fig. \ref{DIC.fig}.

The location of the N-S transition was determined by measuring the
highest nematic volume fraction ($\phi^N$) and the lowest smectic
volume fraction ($\phi^S$) observed. The volume fraction
$\phi^S=c^S v$, where $c^S$ is the number density, and $v$ is the
volume of a single rod $\pi L D^2/4$. A similar equation holds for
$\phi^N$. The concentration of the phases was measured by
absorption spectroscopy with the optical density ($A$) of the
virus being
$A_{\mbox{\scriptsize{269nm}}}^{\mbox{\scriptsize{1mg/ml}}}=3.84$
for a path length of 1 cm.

Since knowing the surface charge of the virus is critical to our
analysis of the N-S transition, we experimentally measured the pH
of the virus solutions at concentrations in the nematic phase just
below the N-S transition. We found that for an initial buffer
solution at pH 8.2 (Tris-HCl buffer pKa$=8.2$), the pH of the
concentrated virus suspensions is slightly less than 8.2, but
still well within the buffering pH range (pH$=$pKa$\pm 1$). The
surface charge does not change significantly over this range
\cite{Zimmermann86}. At pH 5.2 (Acetic acid buffer pKa$=4.76$) the
measured pH of the virus suspensions near the N-S transition was
slightly higher than 5.2, with pH increasing slightly with
decreasing ionic strength, most likely due to the relatively high
concentration of virus counterions (50-100 mM) as compared to
buffer ions (20 mM). This shift further away from the pKa may
influence the phase behavior by increasing the the viral surface
charge. The implications of these measurements are discussed in
the Results section.

\section{Results}
\subsection{Flexibility and ionic strength dependence of the N-S
transition}
\begin{figure}
\centerline{\epsfig{file=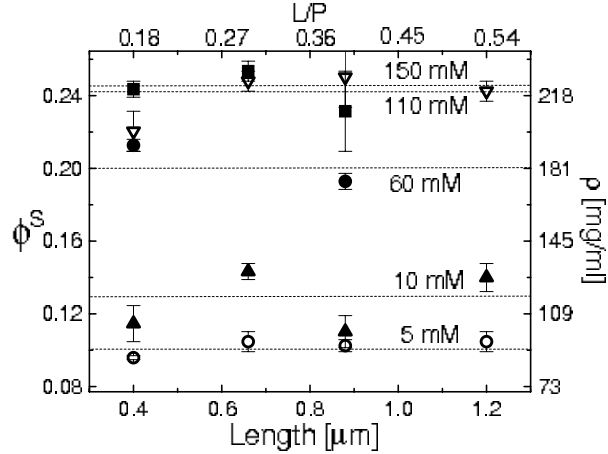,width=8.cm}}\caption[Nematic-Smectic
transition as a function of ionic strength and rod
flexibility]{\label{NSvsL.fig} Volume fraction at the
nematic-smectic phase transition, $\phi^{S}$, for multiple ionic
strengths at pH 8.2 as a function of rod length $L$ and
flexibility $L/P$. On the right axis is the measured concentration
in mass density $\rho^S=\phi^S M /v N_a$, where $N_a$ is
Avagadro's number. Legend for ionic strengths is as follows:
$\circ$ 5 mM, $\blacktriangle$ 10 mM, $\bullet$ 60 mM,
$\bigtriangledown$ 110 mM, $\blacksquare$ 150 mM. With increasing
ionic strength $\phi^{S}$ increases due to increasing
electrostatic screening. Dashed lines are a guide to the eye at
constant ionic strength. Within experimental accuracy the smectic
phase transition is independent of flexibility within the range
$0.18<L/P<0.54$.}
\end{figure}

Fig. \ref{NSvsL.fig} shows $\phi^S$ as a function of the M13
mutant particle length, and therefore virus flexibility by $L/P$,
for multiple ionic strengths. Focusing on how rod flexibility
influences the phase transition, we observe that at each ionic
strength the measured $\phi^{S}$ is independent of virus length,
within experimental accuracy, and thus independent of changing
flexibility in the range of $0.18<L/P<0.54$. The bidispersity of
the 0.64 $\mu$m and 0.39 $\mu$m rod suspensions does not seem to
influence the phase boundary because these samples, which are 20\%
1.2 $\mu$m rods by mass exhibit the same phase behavior as
samples which are 100\% 1.2 $\mu$m rods.

\begin{figure}
\centerline{\epsfig{file=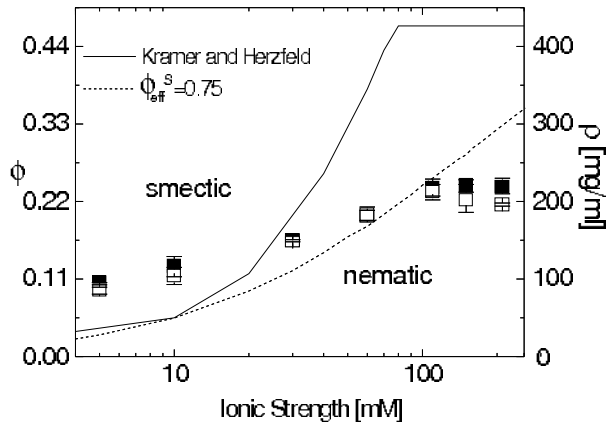,width=8.cm}}
\caption[]{\label{NSave.fig} Average values of $\phi^N$ (open) and
$\phi^S$ (solid) at the N-S transition as a function of ionic
strength at pH 8.2. Average at each ionic strength is over the
results for the four M13 length mutants. The solid line is
$\phi^S$ taken from simulations by Kramer and Herzfeld
\cite{Kramer00} for the N-S transition of particles the same size
as {\it fd} and with a renormalized surface charge of 1e$^-/7.1$
\AA. The dashed line is $\phi^S= \phi^{\mbox{\scriptsize
S}}_{\mbox{\scriptsize eff}}*D^2/(D_{\mbox{\scriptsize
eff}}^{\mbox{\scriptsize N}})^2$ with $\phi^{\mbox{\scriptsize
S}}_{\mbox{\scriptsize eff}}=0.75$. }
\end{figure}

As the N-S phase transition is independent of rod flexibility for
these experiments, we averaged the results for $\phi^S$ and
$\phi^N$ from all particle lengths to study the ionic strength
dependence of the phase transition. These averaged values for
$\phi^S$ and $\phi^N$ are shown as a function of ionic strength in
Fig. \ref{NSave.fig}. We observe that with increasing ionic
strength, and therefore a corresponding decrease in electrostatic
interactions, the volume fraction of the phase transition
increases until an ionic strength of approximately $\mbox{I}=100$~mM, at
which point the phase transition becomes independent of ionic
strength. Whereas the increase in phase transition concentration
with ionic strength has been observed previously for suspensions
of {\it fd} virus \cite{Dogic97}, the plateau in $\phi^S$ and
$\phi^N$ at high ionic strengths is previously undocumented.

\subsection{Surface charge dependence of the N-S transition}
\begin{figure}
\centerline{\epsfig{file=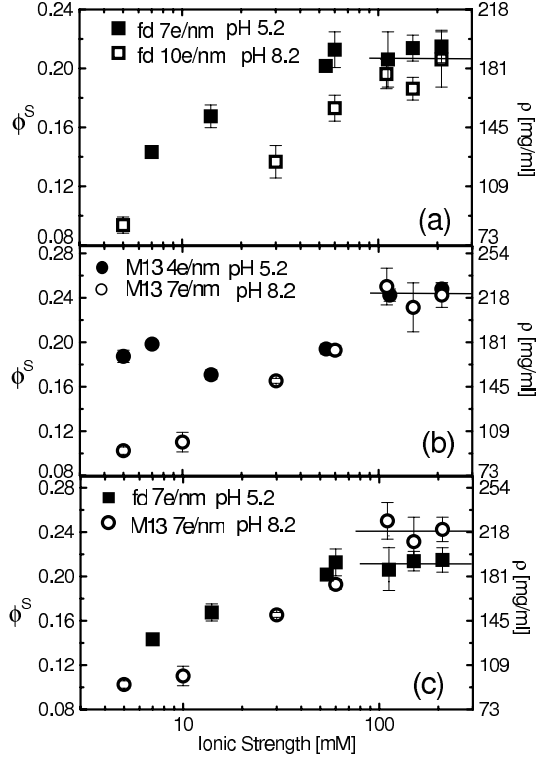,width=7.cm}}
\caption[Nematic-Smectic phase transition as a function of ionic
strength for M13 and {\it fd} at pH 8.2 and 5.2]{\label{NS.fig}
Nematic-smectic phase transition volume fraction $\phi^S$ as
function of ionic strength for suspensions of a) {\it fd} and b)
M13 at pH 5.2 (solid) and pH 8.2 (open). Figure c) shows M13 (pH
8.2) and {\it fd} (pH 5.2) at 7 e/nm surface charge. Solid lines
highlight the ionic strength independence at high ionic strength.}
\end{figure}

To determine the influence of virus surface charge on N-S phase
transition, we measured phase behavior of {\it fd} and M13 at both
pH 8.2 and pH 5.2. Fig. \ref{NS.fig} presents the ionic strength
and pH dependence of the N-S phase transition for {\it fd} (a) and
M13 (b). Below 100 mM, there is a strong pH dependence in the N-S
transition, as shown in Fig. \ref{NS.fig}a,b. Suspensions at
higher pH (higher surface charge) consistently enter a smectic
phase at lower concentrations. Above about 100 mM, $\phi^S$ is
independent of ionic strength, as in Fig. \ref{NSave.fig}. The
{\it fd} phase boundary at high ionic strength saturates around
$\phi_{\mbox{\scriptsize{sat}}}^S\sim 0.21$, independent of pH,
and the M13 phase boundary saturates around
$\phi_{\mbox{\scriptsize{sat}}}^S=0.24$, also independent of pH.
Even at these high ionic strengths, where the surface charge of
the virus is well screened, the higher charged {\it fd}
suspensions have a phase boundary at lower concentrations than the
M13 suspensions.

In Fig. \ref{NS.fig}c we compare the phase transition for M13 and
fd suspensions when both viruses have the same surface charge of
~7e$^-$/nm. Because the rods have the same surface charge, we
assume the rods differ only by the location of the charges
(positive and negative) on the surface. At low ionic strength the
phase behavior is similar, but {\it fd} suspensions consistently
enter the smectic phase at a slightly higher concentration.  We
note that the small measured increase in pH at low ionic strength
of the pH 5.2 viral solutions mentioned in the Materials and
Methods section does not account for this difference. An increase
in pH would lower, not raise, the pH 5.2 phase transition
concentrations by increasing electrostatic interactions. At high
ionic strength, the reverse is true; {\it fd} has a lower phase
transition concentration than M13 suspensions, as mentioned above.
We believe the measured differences in $\phi^S_{\mbox{\scriptsize
sat}}$ between M13 and fd to be statistically significant, and
will discuss this unexpected observation further in the following
section.


\section{Discussion}
\subsection{Flexibility and ionic strength dependence of the N-S
transition} The nematic-smectic transition of flexible, hard rods
has been studied both theoretically and computationally
\cite{Schoot96,Tkachenko96,Polson97}. A small amount of
flexibility is expected to drive the smectic phase to higher
concentrations, from the predicted hard-rigid-rod concentration of
$\phi^S$=0.47 \cite{Bolhuis97a}, to approximately
$0.75\lesssim\phi^{S}\lesssim 0.8$ within the semiflexible limit
\cite{Tkachenko96}. Within the semiflexible limit ($L/P \sim 1$),
however, $\phi^S$ is predicted to be essentially independent of
flexibility \cite{Tkachenko96}. This insensitivity of $\phi^S$ to
flexibility in the semiflexible limit is in agreement with the
measurements presented in Fig. \ref{NSvsL.fig}. We note that this
result is in striking contrast to the significant flexibility
dependence measured at isotropic-nematic transition for this same
system of semiflexible M13 mutants which we describe in a separate
report \cite{Purdy04}.

\begin{figure}
\centerline{\epsfig{file=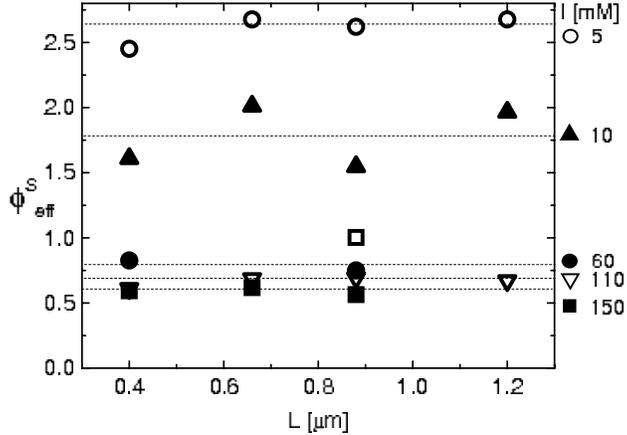,width=8.cm}}\caption[]{\label{phieff.fig}
Effective volume fraction volume fraction along the
nematic-smectic phase transition
$\phi_{\mbox{\scriptsize{eff}}}^S=\phi^S(D^{\mbox{\scriptsize{N}}}_{\mbox{\scriptsize{eff}}})^2/D^2$
for multiple ionic strengths at pH 8.2 as a function of $L$.
$\phi^S$, the actual volume fraction at the N-S transition, is
shown in Fig. \ref{NSvsL.fig}. Legend for symbols is to the right
of the figure. Dashed lines drawn are a guide to the eye and are
at constant ionic strength. Because
$\phi_{\mbox{\scriptsize{eff}}}^S$ strongly depends on ionic
strength, we conclude that
$D^{\mbox{\scriptsize{N}}}_{\mbox{\scriptsize{eff}}}$ does not
describe the electrostatic interactions at high virus
concentrations.}
\end{figure}

As our rods are charged, the ionic strength of the virus
suspension plays an important role in determining the phase
boundaries by screening electrostatic interactions. To compare our
charged-flexible-rod results with current predictions for the N-S
phase transition of hard (rigid or flexible) rods, we have to
effectively account for the electrostatic interactions between our
virus rods. Two methods for incorporating electrostatics into the
N-S phase transition of hard rods
\cite{Stroobants86a,Vroege92,Kramer00} were presented earlier in
this paper. One way to do this is to graph
$\phi_{\mbox{\scriptsize{eff}}}^S$, the measured effective volume
fraction along the nematic-smectic transition, and compare it to
the theoretical volume fraction,
$\phi_{\mbox{\scriptsize{th}}}^S$, for the N-S transition of hard,
semiflexible rods~\cite{Tkachenko96,Polson97}.
$\phi_{\mbox{\scriptsize{eff}}}^S$ is defined as $c^S \pi L
(D_{\mbox{\scriptsize
eff}}^N)^2/4=\phi^S(D^{\mbox{\scriptsize{N}}}_{\mbox{\scriptsize{eff}}})^2/D^2$
and is shown in Fig~\ref{phieff.fig}. If the effect of
electrostatics can be accounted for by replacing $D$ with
$D_{\mbox{\scriptsize eff}}^N$, as can be done at the
isotropic-nematic transition at high ionic strength, we could
predict that $\phi_{\mbox{\scriptsize{eff}}}^S=
\phi_{\mbox{\scriptsize{th}}}^S=0.75$. In other words, if
$D_{\mbox{\scriptsize{eff}}}^N$ accurately models the
interparticle electrostatic interactions, the effective volume
fraction $\phi_{\mbox{\scriptsize eff}}^S$ should be equivalent to
the hard-flexible rod volume fraction and should be independent of
ionic strength. Thus multiplying the measured values for $\phi^S$
shown in Fig. \ref{NSvsL.fig} by
$(D^{\mbox{\scriptsize{N}}}_{\mbox{\scriptsize{eff}}})^2/D^2$
should result in the collapse of all the different ionic strength
data.

However, we find that $\phi_{\mbox{\scriptsize{eff}}}^S$ depends
quite strongly on ionic strength in Fig. \ref{phieff.fig}.
Previously, we observed $\phi_{\mbox{\scriptsize{eff}}}^S=0.75$
independent of ionic strength for suspensions of {\it fd} virus
~\cite{Dogic97}. The data in Fig. \ref{phieff.fig} is consistent
with this value at high ionic strengths ($60
\mbox{mM}<\mbox{I}<150$ mM), but by including a larger range of
ionic strengths as well as multiple particle lengths, we clearly
observe an ionic strength dependence in
$\phi_{\mbox{\scriptsize{eff}}}^N$, with
$\phi_{\mbox{\scriptsize{eff}}}^N$ ranging from 2.5 to 0.5.
Furthermore, the plateau in the phase transition concentration at
high ionic strength is not captured by scaling $\phi^S$ by
$(D^{\mbox{\scriptsize{N}}}_{\mbox{\scriptsize eff}})^2/D^2$, as
shown in Fig. \ref{NSave.fig} by the dashed curve for
$\phi_{\mbox{\scriptsize eff}}^S=0.75$. The large ionic strength
dependence of $\phi_{\mbox{\scriptsize{eff}}}^N$, and the measured
ionic strength independence above 100 mM indicate that
$D^{\mbox{\scriptsize{N}}}_{\mbox{\scriptsize{eff}}}$ is
inadequate for describing the electrostatic interactions at the
N-S transition. This is in contrast to the I-N transition, where
$D_{\mbox{\scriptsize eff}}$ accurately incorporates the
electrostatic interactions between virus rods at high ionic
strength \cite{Purdy04}. Furthermore, because
$\phi_{\mbox{\scriptsize eff}}^S>1$ at low ionic strength we
conclude that $D_{\mbox{\scriptsize eff}}^N$ overestimates the
electrostatic interactions. This is not surprising because
$D^{\mbox{\scriptsize{N}}}_{\mbox{\scriptsize{eff}}}$ is based on
the second virial approximation which, strictly speaking, is valid
only for isotropic suspensions at low concentrations. Using
$D_{\mbox{\scriptsize eff}}$ to relate the phase behavior of
charged rods to hard-rod predictions is inappropriate at the N-S
transition, particularly at high ionic strengths.

The plateau in $\phi^S$ at high ionic strength is indeed predicted
for parallel, charged, rigid spherocylinders with a concentration
dependent avoidance diameter $D_{\mbox{\scriptsize a}}$, as shown
by the solid line in Fig \ref{NSave.fig}. This avoidance model
predicts that at high ionic strength $\phi^S$ saturates at
$\phi_{\mbox{\scriptsize sat}}^S=0.47$, the theoretical value for
hard, rigid spherocylinders ~\cite{Kramer00,Bolhuis97a}. Because
our rods are semiflexible, we would correspondingly predict that
$\phi_{\mbox{\scriptsize sat}}^S$ would be equal to the
theoretical value for hard-semiflexible rods,
$\phi_{\mbox{\scriptsize th}}^S=0.75$. Instead of this value, our
measurements of the phase transition volume fraction as a function
of ionic strength yield
$\phi_{\mbox{\scriptsize{sat}}}^{\mbox{\scriptsize{S}}}=0.21-0.24$,
which is three times lower than predicted by hard flexible rod
theories. This suggests that either the flexibility of the rods
has lowered the phase transition from that of hard rods, in
contradiction to both theories and simulations, or that the
electrostatic interactions between the rods are not accurately
represented by either $D_a$ or $D_{\mbox{\scriptsize eff}}$. We
will return to this question at the end of the Discussion section.

\subsection{Surface charge dependence of the N-S transition}
The pH dependence of the phase transition visible in Fig.
\ref{NS.fig}, and the difference between M13 and {\it fd} saturation concentrations present even when they share the same
surface charge (Fig. \ref{NS.fig}c) also indicate that neither
$D_{\mbox{\scriptsize eff}}$ nor $D_a$ are appropriate for
describing the electrostatic interactions between the virus rods
at the N-S transition. The non-linearity of the Poisson-Boltzmann
equation predicts that for high linear charge density the
long-range electrostatic potential between rods is insensitive to
surface charge changes and thus pH changes
\cite{Onsager49,Stroobants86}. This is confirmed at the
isotropic-nematic transition, where the charge dependence is well
described by $D_{\mbox{\scriptsize eff}}$ and the pH dependence of
the phase transition is very small \cite{Purdy04}. However, a
strong pH dependence of the smectic phase transition at low ionic
strengths for both {\it fd} (Fig. \ref{NS.fig}a) and M13 (Fig.
\ref{NS.fig}b) suspensions is observed. The observed difference
between M13 and {\it fd} saturation concentrations at high ionic
strength and equal surface charge (Fig. \ref{NS.fig}c) is also not
expected from Poisson-Boltzmann theory. This high sensitivity of
the N-S transition to changes in pH and surface charge
configuration indicates that the charge independent nature
predicted by both $D_{\mbox{\scriptsize eff}}$ and $D_a$ does not
correctly characterize the electrostatic interactions at the
concentrations of the N-S transition.

 One possible explanation for why the high ionic strength,
and correspondingly high concentration, phase behavior is
sensitive to surface charge configuration is that the adjacent
virus surfaces are separated by approximately one virus diameter
(6.6 nm) \cite{Purdy03}, which is on the order of the spacing
between viral coat-proteins (1.6 nm), and the Debye screening
length $\kappa=3.0\mbox{\AA}/\sqrt{I}=9$\AA. When the surface to
surface distance is of the order of the Debye screening length,
the continuous charge distribution approximation used in Poisson
Boltzmann theory can no longer be used, as is done in the
effective diameter calculations. Furthermore, it has been shown
theoretically that discretization of the surface charges can
change the predicted counterion condensation from that predicted
by the non-linear Poisson Boltzmann equation \cite{Marzec94,
Henle04}. Perhaps it is because we are in the regime where the
surface charge configuration can no longer be neglected that we
observe charge-configuration-dependent saturation of the
nematic-smectic phase transition. Previous work by Lyubartsev {\it
et. al.} has been done to simulate the electrostatic interactions
between these virus rods in the presence of divalent ions using an
approximate discrete charge configuration \cite{Lyubartsev98}. We
propose that theoretical models or simulations similar to those by
Lyubartsev {\it et. al} of the electrostatic interactions of a
dense, rod-like polyelectrolyte system which include the detail of
the surface charge configuration including the location of
positive and negative amino acids on the viral surface, may shed
light on the experimental differences between M13 and {\it fd}
nematic-smectic transitions at high ionic strength.

\subsection{Origin of the ionic strength independence of the NS transition}
At high ionic strength we have measured an ionic strength
independent N-S transition. This is similar to the phase behavior
predicted by Kramer and Herzfeld, in which the phase transition
volume fraction saturates at that predicted for the N-S transition
of  hard rods.  Yet, the value for our measured
$\phi_{\mbox{\scriptsize sat}}^S$ is three times lower than the
predicted N-S transition volume fraction for semi-flexible rods
$\phi_{\mbox{\scriptsize th}}^S=0.75$. These observations point to a failure of
theory to describe the role of electrostatics and/or flexibility
on the N-S transition.

It has been shown by Odijk that undulations of semiflexible rods
in a hexagonal configuration are typically contained within a tube
of a diameter larger than the bare rod diameter \cite{Odijk86}.
These undulations create an effective repulsion similar to
Helfrich repulsion of membranes. The dominant interaction between
charged flexible rods in a dense suspension therefore depends on
the relative size of the tube diameter and the electrostatic
effective diameter \cite{Odijk93}. At low ionic strength the
interparticle interactions would be dominated by electrostatics,
and at high ionic strength the interparticle interactions would be
dominated by steric interactions of the flexible rods. This simple
argument qualitatively agrees with our observed phase behavior. If
charged, flexible rod behavior is indeed dominated by steric
interactions at high ionic strengths, we expect that the N-S phase
transition would be at a value equal to that predicted by theory
for hard-semiflexible rods. However, both theory and simulations
for hard-flexible rods which incorporate this repulsion due to
flexibility, predict that flexibility destabilizes the N-S
transition, subsequently increasing, not decreasing as measured,
the transition concentration above that predicted for rigid rods
\cite{Tkachenko96,Polson97,Hidalgo05}. It is possible that the
role of flexibility is not accurately incorporated into the
theories which predict an increase in $\phi^S$ from that predicted
for rigid rods. However, we believe these theories to be accurate,
specifically because simulations by Polson and Frenkel of hard
semiflexible rods demonstrate that flexibility increases the
N-S transition concentration above that predicted for rigid rods
\cite{Polson97}.

A second possible explanation for the discrepancy between our
experimental values of $\phi_{\mbox{\scriptsize sat}}^S$ and the
predicted values for the phase transition of semiflexible hard
rods is that the electrostatic interactions between the rods are
indeed significant at high ionic strength, and that they are
different from the predicted electrostatic interactions. As our
measurements suggest that both the second virial and avoidance
approximations for the electrostatic interactions describe a N-S
transition which differs significantly from the experimentally
observed behavior, we believe this is possible. Further evidence
to suggest that electrostatics still plays a role at high ionic
strength is visible in Fig. \ref{NS.fig} which shows that the
surface charge of the rods can still influence the phase behavior,
even when the phase transition has become ionic strength
independent.

A third possibility is that coupling electrostatics and
flexibility produces an inter-rod repulsion which is a complex
combination of flexible-hard rod and charged-rigid rod
interactions. It has been shown for concentrated suspensions of
DNA, that fluctuations due to the flexibility of the DNA actually
enhance inter-rod repulsions in an exponential manner
\cite{Strey97}. The consequence is that for a given osmotic
pressure exerted by a concentrated DNA suspension, the volume
fraction of DNA is much lower than predicted by Poisson-Boltzman
electrostatics alone. This hypothesis is consistent with our
measurements of a $\phi^S_{\mbox{\scriptsize sat}}$ that is much
lower than predicted for both rigid and flexible rods.

As our results are quite unexpected with respect to current
theoretical predictions for the nematic - smectic transition of rod suspensions,
we have speculated as to possible explanations for our observations. To
fully understand the phase behavior of charged, flexible rods
further computational and theoretical work is clearly needed.

\section{Conclusion}
We have examined the nematic-smectic phase diagram for charged,
semiflexible virus rods as a function of length, surface charge
and ionic strength. We observed that in the semiflexible-rod limit
the N-S phase boundary is independent of rod flexibility, as
predicted theoretically. However, by studying the ionic strength
dependance of this transition we observed that renormalizing the
measured phase transition volume fraction, $\phi^S$, by Onsager's
effective diameter, $D_{\mbox{\scriptsize{eff}}}$, does not
produce an ionic-strength independent phase transition
concentration. Therefore the second virial approximation cannot be
used to map the measured nematic - smectic phase transition of
charged, flexible rods onto hard, flexible rod theories.

At high ionic strength, we found that the concentration of the N-S
phase boundary is independent of ionic strength. Kramer and
Herzfeld's avoidance diameter theory~\cite{Kramer00} qualitatively
reproduces the observed ionic strength independent N-S phase
behavior, but predicts that the ionic strength independent phase
boundary is equal to the predicted hard-rod phase boundary. Our
experimental results, however, are three times lower than the
hard-particle phase boundary predicted for semiflexible rods,
$\phi_{\mbox{\scriptsize{th}}}^S$~\cite{Tkachenko96,Polson97}.
Clearly, more theoretical work is needed to understand the
nematic-smectic phase transition of charged, flexible rods, in
order to reconcile the differences we observe with charged,
flexible viruses and that reported in simulations of hard,
flexible rods.

Finally, significant differences were measured between M13 and
{\it fd} nematic-smectic phase transition concentrations, even when they shared
the same average surface charge. These results indicate that the
electrostatic interactions between these rods are more
complicated than can be accounted for by calculating the
interparticle potential assuming a uniform renormalized surface
charge. We hypothesize that the electrostatic interactions between
rods is influenced by the configuration of the charged amino acids
on the viral surface. Experimental tests of this hypothesis could be
made by measuring M13 and {\it fd} equations of state (pressure
vs density), and thus the particle-particle interactions, as a
function of solution salt and pH, as in techniques developed for
DNA \cite{Strey97}. Computationally, this hypothesis can be tested
by calculating the pair potential between rods with discrete
charges \cite{Lyubartsev98}.

\begin{acknowledgments}
We acknowledge support from the NSF(DMR-0088008).
\end{acknowledgments}

\end{document}